%

\magnification 1000

\hsize= 16 truecm
\vsize= 22 truecm

\hoffset=-0.0 truecm
\voffset= +1 truecm

\baselineskip=14 pt

\parindent 15pt 

\footline={\iftitlepage{\hfil}\else
       \hss\tenrm-- \folio\ --\hss\fi}	

\parskip 0pt 


\font\bello= cmr10 scaled \magstep2
\font\piccolo=cmr8

\font\pbf=cmbx8

\def\eq{\autoeqno}
\def\re{\eqrefp}


\def\eq#1{\autoeqno{#1}}
\def\re#1{\eqrefp{#1}}

\newcount\notenumber \notenumber=1
\def\nota#1{\unskip\footnote{$^{\the\notenumber}$}{\piccolo #1}%
  \global\advance\notenumber by 1}

\def\s{\scriptstyle}  

\catcode`@=11 
%

\newcount\cit@num\global\cit@num=0

\newwrite\file@bibliografia
\newif\if@bibliografia
\@bibliografiafalse

\def\lp@cite{[}
\def\rp@cite{]}
\def\trap@cite#1{\lp@cite #1\rp@cite}
\def\lp@bibl{[}
\def\rp@bibl{]}
\def\trap@bibl#1{\lp@bibl #1\rp@bibl}

\def\refe@renza#1{\if@bibliografia\immediate        
    \write\file@bibliografia{
    \string\item{\trap@bibl{\cref{#1}}}\string
    \bibl@ref{#1}\string\bibl@skip}\fi}

\def\ref@ridefinita#1{\if@bibliografia\immediate\write\file@bibliografia{ 
    \string\item{?? \trap@bibl{\cref{#1}}} ??? tentativo di ridefinire la 
      citazione #1 !!! \string\bibl@skip}\fi}

\def\bibl@ref#1{\se@indefinito{@ref@#1}\immediate
    \write16{ ??? biblitem #1 indefinito !!!}\expandafter\xdef
    \csname @ref@#1\endcsname{ ??}\fi\csname @ref@#1\endcsname}

\def\c@label#1{\global\advance\cit@num by 1\xdef            
   \la@citazione{\the\cit@num}\expandafter
   \xdef\csname @c@#1\endcsname{\la@citazione}}

\def\bibl@skip{\vskip 0truept}


\def\stileincite#1#2{\global\def\lp@cite{#1}\global
    \def\rp@cite{#2}}
\def\stileinbibl#1#2{\global\def\lp@bibl{#1}\global
    \def\rp@bibl{#2}}

\def\citpreset#1{\global\cit@num=#1
    \immediate\write16{ !!! cit-preset = #1 }    }

\def\autobibliografia{\global\@bibliografiatrue\immediate
    \write16{ !!! Genera il file \jobname.BIB}\immediate
    \openout\file@bibliografia=\jobname.bib}

\def\cref#1{\se@indefinito                  
   {@c@#1}\c@label{#1}\refe@renza{#1}\fi\csname @c@#1\endcsname}

\def\cite#1{\trap@cite{\cref{#1}}}                  
\def\ccite#1#2{\trap@cite{\cref{#1},\cref{#2}}}     
\def\cccite#1#2#3{\trap@cite{\cref{#1},\cref{#2},\cref{#3}}}
\def\ccccite#1#2#3#4{\trap@cite{\cref{#1},\cref{#2},\cref{#3},\cref{#4}}}
\def\ncite#1#2{\trap@cite{\cref{#1}--\cref{#2}}}    
\def\upcite#1{$^{\,\trap@cite{\cref{#1}}}$}               
\def\upccite#1#2{$^{\,\trap@cite{\cref{#1},\cref{#2}}}$}  
\def\upncite#1#2{$^{\,\trap@cite{\cref{#1}-\cref{#2}}}$}  

\def\clabel#1{\se@indefinito{@c@#1}\c@label           
    {#1}\refe@renza{#1}\else\c@label{#1}\ref@ridefinita{#1}\fi}

\def\biblskip#1{\def\bibl@skip{\vskip #1}}           

\def\insertbibliografia{\if@bibliografia             
    \immediate\write\file@bibliografia{ }
    \immediate\closeout\file@bibliografia
    \catcode`@=11\input\jobname.bib\catcode`@=12\fi}


\def\commento#1{\relax} 
\def\biblitem#1#2\par{\expandafter\xdef\csname @ref@#1\endcsname{#2}}



%
%
\def\b@lank{ }


\newif\iftitlepage      \titlepagetrue

\def\titoli#1{
         \xdef\prima@riga{#1}\voffset+20pt
        \headline={\ifnum\pageno=1
             {\hfil}\else\hfil{\piccolo \prima@riga}\hfil\fi}}

\def\duetitoli#1#2{
                    \voffset=+20pt
                    \headline={\iftitlepage{\hfil}\else
                              {\ifodd\pageno\hfil{\piccolo #2}\hfil
             \else\hfil{\piccolo #1}\hfil\fi}\fi} }

\def\la@sezionecorrente{0}

\catcode`@=12


\autobibliografia

%
\catcode`@=11 
%
%
\def\b@lank{ }

\newif\if@simboli
\newif\if@riferimenti
\newif\if@bozze

\newwrite\file@simboli
\def\simboli{
    \immediate\write16{ !!! Genera il file \jobname.SMB }
    \@simbolitrue\immediate\openout\file@simboli=\jobname.smb}

\def\bozze{\@bozzetrue}

\newcount\eq@num\global\eq@num=0
\newcount\sect@num\global\sect@num=0

\newif\if@ndoppia
\def\numerazionedoppia{\@ndoppiatrue\gdef\la@sezionecorrente{\the\sect@num}}

\def\se@indefinito#1{\expandafter\ifx\csname#1\endcsname\relax}
\def\spo@glia#1>{} 

\newif\if@primasezione
\@primasezionetrue

\def\s@ection#1\par{\immediate
    \write16{#1}\if@primasezione\global\@primasezionefalse\else\goodbreak
    \vskip\spaziosoprasez\fi\noindent
    {\bf#1}\nobreak\vskip\spaziosottosez\nobreak\noindent}
%

\def\sezpreset#1{\global\sect@num=#1
    \immediate\write16{ !!! sez-preset = #1 }   }

\def\spaziosoprasez{26pt plus5pt minus3pt}
\def\spaziosottosez{15pt}

\def\sref#1{\se@indefinito{@s@#1}\immediate\write16{ ??? \string\sref{#1}
    non definita !!!}
    \expandafter\xdef\csname @s@#1\endcsname{??}\fi\csname @s@#1\endcsname}

\def\autosez#1#2\par{
    \global\advance\sect@num by 1\if@ndoppia\global\eq@num=0\fi
    \xdef\la@sezionecorrente{\the\sect@num}
    \def\usa@getta{1}\se@indefinito{@s@#1}\def\usa@getta{2}\fi
    \expandafter\ifx\csname @s@#1\endcsname\la@sezionecorrente\def
    \usa@getta{2}\fi
    \ifodd\usa@getta\immediate\write16
      { ??? possibili riferimenti errati a \string\sref{#1} !!!}\fi
    \expandafter\xdef\csname @s@#1\endcsname{\la@sezionecorrente}
    \immediate\write16{\la@sezionecorrente. #2}
    \if@simboli
      \immediate\write\file@simboli{ }\immediate\write\file@simboli{ }
      \immediate\write\file@simboli{  Sezione 
                                  \la@sezionecorrente :   sref.   #1}
      \immediate\write\file@simboli{ } \fi
    \if@riferimenti
      \immediate\write\file@ausiliario{\string\expandafter\string\edef
      \string\csname\b@lank @s@#1\string\endcsname{\la@sezionecorrente}}\fi
    \goodbreak\vskip 48pt plus 60pt
    \noindent\if@bozze\llap{\it#1\quad }\fi
      {\bf\the\sect@num.\quad #2}\par\nobreak\vskip 15pt
    \nobreak\noindent}

\def\semiautosez#1#2\par{
    \gdef\la@sezionecorrente{#1}\if@ndoppia\global\eq@num=0\fi
    \if@simboli
      \immediate\write\file@simboli{ }\immediate\write\file@simboli{ }
      \immediate\write\file@simboli{  Sezione ** : sref.
          \expandafter\spo@glia\meaning\la@sezionecorrente}
      \immediate\write\file@simboli{ }\fi
    \s@ection#2\par}


\def\eqpreset#1{\global\eq@num=#1
     \immediate\write16{ !!! eq-preset = #1 }     }

\def\eqref#1{\se@indefinito{@eq@#1}
    \immediate\write16{ ??? \string\eqref{#1} non definita !!!}
    \expandafter\xdef\csname @eq@#1\endcsname{??}
    \fi\csname @eq@#1\endcsname}

\def\eqlabel#1{\global\advance\eq@num by 1
    \if@ndoppia\xdef\il@numero{\la@sezionecorrente.\the\eq@num}
       \else\xdef\il@numero{\the\eq@num}\fi
    \def\usa@getta{1}\se@indefinito{@eq@#1}\def\usa@getta{2}\fi
    \expandafter\ifx\csname @eq@#1\endcsname\il@numero\def\usa@getta{2}\fi
    \ifodd\usa@getta\immediate\write16
       { ??? possibili riferimenti errati a \string\eqref{#1} !!!}\fi
    \expandafter\xdef\csname @eq@#1\endcsname{\il@numero}
    \if@ndoppia
       \def\usa@getta{\expandafter\spo@glia\meaning
       \la@sezionecorrente.\the\eq@num}
       \else\def\usa@getta{\the\eq@num}\fi
    \if@simboli
       \immediate\write\file@simboli{  Equazione 
            \usa@getta :  eqref.   #1}\fi
    \if@riferimenti
       \immediate\write\file@ausiliario{\string\expandafter\string\edef
       \string\csname\b@lank @eq@#1\string\endcsname{\usa@getta}}\fi}

\def\autoreqno#1{\eqlabel{#1}\eqno(\csname @eq@#1\endcsname)
       \if@bozze\rlap{\it\quad #1}\fi}
\def\autoleqno#1{\eqlabel{#1}\leqno\if@bozze\llap{\it#1\quad}
       \fi(\csname @eq@#1\endcsname)}
\def\eqrefp#1{(\eqref{#1})}
\def\numeriadestra{\let\autoeqno=\autoreqno}
\def\numeriasinistra{\let\autoeqno=\autoleqno}
\numeriadestra

\catcode`@=12

\numerazionedoppia

\def\bbuildrel#1_#2{\mathrel{\mathop{\kern 0pt#1}\limits_{#2}}}


\vglue 3 truecm
{\parindent 0 pt

{\bello     An investigation of the hidden structure of states in a 
            mean field spin glass model.                            }

\vskip 1 truecm
Andrea Cavagna, Irene Giardina and Giorgio Parisi
\vskip 0.5 truecm
{\it          Dipartimento di Fisica,
Universit\`a di Roma I `La Sapienza',
P.le A. Moro 5, 00185 Roma, Italy 

INFN Sezione di Roma I, Roma, Italy

\vskip 0.5 truecm                  }

{\sl cavagna@roma1.infn.it}

{\sl giardina@roma1.infn.it}

{\sl parisi@roma1.infn.it}

\vskip 1 truecm 

June 1997            }
 
\vskip 2 truecm 

\noindent
{\bf Abstract.}
\vskip 0.5 truecm

We study the geometrical structure of the states in the low
temperature phase of a mean field model for generalized spin glasses,
the $p$-spin spherical model. This structure cannot be revealed by
the standard methods, mainly due to the presence of an exponentially
high number of states, each one having a vanishing weight in the
thermodynamic limit. Performing a purely entropic computation, based
on the TAP equations for this model, we define a constrained
complexity which gives the overlap distribution of the states. 
We find that this distribution is continuous, non-random and highly 
dependent on the energy range of the considered states.
Furthermore, we show which is the geometrical
shape of the threshold landscape, giving some insight into the role 
played by threshold states in the dynamical behaviour of the system.


\vskip 1 truecm


\vfill\eject
\titlepagefalse


\autosez{intro} Introduction.
\par

Despite some recent developments \cccite{kpz}{franzparisi}{noi},  a
deep understanding of the structure of the states in the $p$-spin
spherical model is still lacking.  The problem is the following: 

In the context of the TAP approach \cite{tap}, it has been shown that,
in the temperature range between the static and the dynamical
transition,  this model has an exponentially high number of states
(TAP solutions), with free energy densities in a finite range
$[f_{min},f_{th}]$. What happens is that {\it equilibrium}
states at temperature $T$ are not the lowest ones corresponding to
$f_{min}$,  but rather those which optimize the balance between the
free energy and  the entropic contribution due to the presence of a
great number of  states with the same free energy. Thus equilibrium
states are those   which minimize the generalized free energy density
$\phi(f)=f-T\Sigma(f)$, where $\Sigma$ is the {\it complexity}.  The
states with free energy density $f$ either lower or larger then the
value which minimizes $\phi$ must be considered as metastable.  On the
other hand, {\it all} these states, the equilibrium as well  as the
metastable ones, singularly taken have a vanishing weight in the
thermodynamic limit. In this sense, an equilibrium state is not
different from a metastable state, since the equilibrium condition is
a fully collective effect \ccite{kpz}{crisatap}. 

It is clear that the presence of this huge number of states makes it
interesting to know how they are disposed in the phase space,
and therefore to investigate their distribution and structure.

To clarify what we intend with structure of the states it is useful to
think about the Sherrington-Kirkpatrick (SK) model \cite{sk}.  In this
case, in the context of the replica method,  the solution given in
\cccite{rsb1}{rsb2}{rsb3} has permitted to define and calculate the
overlap distribution $P(q)$ of  the pure states: $P(q)$ is defined as
the probability   that, picked up two states, their mutual overlap is
equal  to $q$. Therefore, in the distribution $P(q)$ two different
contributions are present:  the existence of states having mutual
overlap $q$  and their individual statistical weights.  The function
$P(q)$ gives for the SK model an important structural information on
the distribution of the states in term of the overlap \cite{spin}. 

For the $p$-spin spherical model we would like to have a structural
information of the same kind as the one  given by the distribution
$P(q)$ for the SK model.

Unfortunately, it is known that applying the standard replica method
to the $p$-spin spherical model a trivial result is obtained
\cite{crisa1}:  in the intermediate temperature phase that we are
considering,  the model is solved by a replica symmetric solution,
corresponding to a trivial distribution function $P(q)=\delta(q)$.
This delta function   simply means that the {\it typical} overlap
between  two states is zero. On the other hand, one can wonder why
this distribution does not get contribution from the self overlap of
the equilibrium states, which is different from zero.  The answer is
that all these states have singularly a weight so low that the
contribution of their self overlap is not present in the distribution
$P(q)$.  In other words, it is highly unlikely to pick up twice  the
same state and measure its self overlap, either if this is an
equilibrium state or a metastable one.

For the same reason it is possible that the distribution $P(q)$ does
not catch a contribution from {\it all}  the other possible values of
the overlap $q$,  simply because picking up two states with overlap
different from zero has a vanishing probability in the thermodynamic
limit.  This would mean that the trivial form of $P(q)$ is not a
consequence of the absence of states with mutual overlap different
from zero, but rather  of the difficulty of finding them \cite{vira}.  
On the contrary, there is the possibility that indeed {\it all} the states of
this model have mutual overlap zero, i.e. that states with overlap
different from zero do not exist at all. In this last case it is clear
that there would be only a trivial structure of the states, exactly
reproduced by a trivial $P(q)$.  The standard static approaches cannot
distinguish between these two pictures, and more than this, in the
case in which a non trivial hidden structure were present, they are
not able to give us any insight into the problem.  In our opinion
it seemed very strange to have this huge number of states  without any
interesting geometrical structure and therefore we have tried  to
develop some non standard methods which could provide some information
on this topic.

The first question  to answer is then if a non trivial structure of
the  states,  worth being investigated, exists.  This question has
been partially answered in the context of the  real replica method, by
the definition of  a three replica potential \cite{noi}. Within this
method it has been possible  to demonstrate  that, given an
equilibrium state, there are other states, both metastable and
equilibrium,  at {\it any} value of the overlap $q$ with it, until a
certain maximum value.  This shows clearly that a non trivial
structure of the states for this model is present,  and that the second
picture we have stated  above has to be discarded. On the other hand,
the shape of the  energy spectrum of the states found with this method
was not completely understood and besides there was no control on
the choosing procedure of  the detected states.

Therefore it is necessary to define a tool by which the hidden
structure of the states for the $p$-spin spherical  model can be
analyzed in a deep way.  Yet, as we have seen, there is the problem of
the vanishing weight of these  states, that leads to the trivial form
of the standard distribution function $P(q)$. 

Bearing this in mind, the most natural thing to do is to perform a
purely entropic computation, disregarding the thermodynamic weight of
the states.  Thus, what we have done is to fix a reference state in
the phase space and simply {\it count} how many states of a given kind
are  present at overlap $q$ with it. More precisely, what we have
actually  computed is the number of TAP solutions having a given
overlap $q$ with a single fixed solution. The resulting quantity is what we
have called the {\it constrained complexity} $\Sigma_c$.

To conclude this Introduction we want to stress a point that could
seem  trivial, but that has some  importance in our opinion. We said
that we wanted to study the structure of the {\it states} of this
model, but actually we work with   {\it solutions} of the TAP
equations.  The underlying hypothesis is then that TAP solutions
really correspond to thermodynamic states, intended as local minima of
the {\it true} free energy of the system. This is not a trivial
identification, but it has been confirmed in various ways 
\cccite{franzparisi}{buribarrameza}{noi}. For example, in the case of the three
replica potential of \cite{noi}, one can show that the local minima of
the potential, which correspond to metastable states of the system,
always have a free energy and a self overlap that satisfy TAP
equations.

The Paper is organized in the following way: In Section 2 we 
define the constrained complexity and describe the way in which the
calculation has been performed.  The main results are exposed in
Section 3, where the behaviour of $\Sigma_c$  is analyzed and
interpreted in terms of geometrical structure of the states.  In
Section 4 we address the question of which are the dominant states at
a certain distance from a reference equilibrium state,  while in
Section 5 we focus on the structure of the threshold states, 
which are  important under many respects. In Section 6
we state the conclusions and outline the most important open problems. 
Finally, the comparison with the results of the real replica method
is carried out in a detailed way in Appendix A.


\autosez{ilcalcolo} The constrained complexity.
\par

The $p$-spin spherical model is defined by the following Hamiltonian
$$
H(s)=-\!\!\!\!\sum_{i_1<\dots <i_p} J_{i_1\dots i_p} s_{i_1}\dots s_{i_p}
\eq{mod}
$$
where the spins $s$ are real variables satisfying the spherical
constraint $\sum_i s_i^2 = N$ ($N$ is the size of the system) and the
couplings  $J$ are Gaussian variables with zero mean and variance
$p!/2N^{p-1}$
\cccite{grome}{tirumma}{crisa1}. 

In the frame of the TAP approach \cite{tap}, one formulates mean field
equations for the local magnetizations $m_i=\langle s_i\rangle$ of the
system.  In \cite{kpz} it has been introduced a free energy density
$f_{TAP}$, function of the magnetizations $m_i$, minimizing which one
obtains the TAP equations of the system. Solving these equations at
$T=0$ one finds  minima of $f_{TAP}$ with energy density included in a
finite range $[E_{min},E_{th}]$. The solutions with the highest energy
density $E_{th}$ are usually called {\it threshold}  solutions.  For
each value of the energy density $E$ in this range, there is an
exponentially high number of solutions
$$ 
{\cal N} (E) \sim e^{N \Sigma (E)}
\eq{siga}
$$
where $\Sigma(E)$ is the {\it complexity} of the class of TAP
solutions  corresponding to that particular energy. The complexity
$\Sigma(E)$ for this  model has been  computed in \cite{crisatap},
where it is found that $\Sigma$ is an increasing  function of $E$,
which is zero for $E=E_{min}$ and reaches a finite  value for
$E=E_{th}$.  Moreover, due to the particular homogeneity of the
Hamiltonian, there is a one to one mapping of the  solutions found at
temperature zero,  into solutions at finite temperature $T$, without
splitting nor merging of  solutions with varying the temperature.  Due
to this, one can solve TAP equations at $T=0$,  obtaining a class of
solutions with a certain zero temperature energy density $E$ and then
transport  these solutions to   finite $T$. Therefore from now on we
will identify a TAP solution with its  zero temperature energy density
$E$, even if we are considering this solution  at finite temperature
$T$. The important thing is that the complexity  $\Sigma(E)$ of a
given class does not depend on $T$, but only on the zero  temperature
energy $E$ of this class, while the self overlap of each solution
depends both on $E$ and on the temperature $T$.

We now introduce the constrained complexity:
$$
\Sigma_c(E_2,q|E_1)
\ \buildrel\hbox{\sevenrm def}\over=
\lim_{N\to \infty}{1\over N}
\ \overline{\log {\cal N}(E_2,q|E_1) }  \ .
\eq{lei}
$$
In this definition ${\cal N}(E_2|q,E_1)$ is the number of TAP
solutions  with energy density $E_2$ having overlap $q$ with a single
fixed solution  of energy density   $E_1$. The bar indicates the
average over the disorder. What we are doing here is to fix a single
state of energy $E_1$ and simply count how many states of energy $E_2$
are found at overlap (distance) $q$ with it. We remark that $q$ is
the overlap between these TAP solutions at finite temperature, while $E_1$ and
$E_2$ are their zero temperature energy densities.

In definition
\re{lei} we have averaged the logarithm of $\cal N$, since we expect that this 
is the extensive quantity. Therefore, to perform this average it is
necessary to introduce replicas. However, it can be shown that in the
unconstrained case 
\cite{crisatap}, the correct ansatz for the overlap matrix is symmetric and 
diagonal and this is equivalent to average directly the number $\cal
N$ of  the solutions. In our case the same prescription leads to the
following  definition:
$$
\Sigma_c(E_2,q|E_1)
\ \buildrel\hbox{\sevenrm def}\over=
\ \lim_{N\to \infty}{1\over N}
\log\overline{ {\cal N}(E_2,q|E_1) } 
\eq{lei2}
$$ 
and this is the quantity we shall compute.  It is surely  possible
that the introduction of the constraint $q$ requires a breaking of the
replica symmetry and therefore definition \re{lei2} has to be
considered as a first approximation. Yet, as we shall show, the
results obtained with  formula \re{lei2} suggest that this is a good
approximation.

The complexity $\Sigma_c$ can be obtained by counting all pairs of
solutions  with energies $(E_1,E_2)$ at mutual overlap $q$ and
dividing it by the number of solutions with energy $E_1$, i.e.
$$
\Sigma_c(E_2,q|E_1)=  \lim_{N\to \infty}\left\{
{1\over N}\log \overline{ {\cal N}(E_1,E_2,q) } -{1\over N}\log
\overline{ {\cal N}(E_1) }\right\}    
\ \buildrel\hbox{\sevenrm def}\over= \Gamma(E_1,E_2,q) - \Sigma(E_1)   
\eq{def}
$$
where $\Sigma(E_1)$ is the usual unconstrained complexity computed in 
\cite{crisatap}. 
We note that the quantity $\Gamma$ is symmetric in $(E_1,E_2)$  while
$\Sigma_c$ is not.  Moreover,
$\Gamma(E_1,E_2,0)=\Sigma(E_1)+\Sigma(E_2)$, since {\it almost}  all
TAP solutions are mutually orthogonal, and then 
$$
\Sigma_c(E_2,0|E_1)=\Sigma(E_2) \ .
\eq{check}
$$
This is the first check we have to face in our calculation.

Let $m^{(1)}_i$ and   $m^{(2)}_i$  be two solutions with self overlaps
$q_1$ and $q_2$, and mutual overlap $q$;  using the notation $m\cdot
m'=\sum_i m_i m'_i$, we have
$$
m^{(1)}\cdot m^{(1)}=N q_1 \quad ; \quad m^{(2)}\cdot m^{(2)}=N q_2
\quad ;
\quad m^{(1)}\cdot m^{(2)}=N q
\eq{ove}
$$
Following \cite{kpz} we write TAP equations in terms of the angular
part of the magnetizations $m_i$
$$
\sigma_i ={ m^{(1)}_i \over   \sqrt{q_1}}   \quad ; \quad   \tau_i = {
m^{(2)}_i \over   \sqrt{q_2}} 
\eq{tra}
$$
for which it holds
$$
\sigma \cdot \sigma = N \quad ; \quad \tau \cdot \tau = N \quad  ; \quad
\sigma \cdot \tau =N { q\over \sqrt{q_1 q_2}}
\buildrel\hbox{\sevenrm def}\over= N q_0 \ .
\eq{sig}
$$
In terms of the angular variables $\sigma$ and $\tau$, the TAP
equations read 
$$
0 = - p\!\!\sum_{i_2<\dots <i_p}J_{i,i_2\dots i_p}\,
\sigma_{i_2}\dots\sigma_{i_p} - pE_1\sigma_i
\ \buildrel\hbox{\sevenrm def}\over= {\cal T}_i(\sigma;E_1) 
\quad,\quad i=1,\dots,N
\eq{tap}
$$
where $E_1$ is the zero temperature energy density
$$
E_1=-{1\over N}\!\sum_{i_1<\dots<i_p}J_{i_1\dots i_p}
\sigma_{i_1}\dots \sigma_{i_p}  \ .
\eq{ene}
$$
Relations of the same kind of \re{tap} and \re{ene} hold for $\tau$
and $E_2$.  It is now possible to write  $\Gamma$ with the standard
method of \ccite{braymoore}{crisatap} in the  following way:
$$
\eqalign{
& \Gamma (E_1,E_2,q)  = \cr & = {1\over N} \log \int {\cal  D}P(J)
\int  {\cal  D}\sigma {\cal  D}\tau  \prod_i\ \delta({\cal
T}_i(\sigma;E_1)) \ \delta({\cal T}_i(\tau;E_2)) 
\ \left| 
\det\left({\partial {\cal T}(\sigma;E_1)\over \partial \sigma} \right)
\right|
\ \left|
\det\left({\partial {\cal T}(\tau;E_2)\over \partial \tau} \right)
\right|
\times
\cr &
\times
\delta(\sigma
\cdot
\sigma
-N) \
\delta(\tau
\cdot
\tau
-N) \
\delta(\sigma
\cdot
\tau
-Nq_0)
\cr}
\eq{mink}
$$
with
$$
{\cal D}P(J)=\prod_{i_1<\dots <i_p}\sqrt{N^{p-1}\over \pi p!}
\exp(-J^2_{i_1\dots i_p}N^{p-1}/p!)\ dJ_{i_1\dots i_p} \ .
\eq{gei}
$$
We note that the dependence on the temperature is entirely contained
in $q_0$ through $q_1$ and $q_2$, functions respectively of $E_1,E_2$
and $\beta$ 
\cite{kpz}.

In formula \re{mink} we can drop the two modulus since it is possible
to check {\it a posteriori} that the determinants are positive.
Actually, this is a tricky point. As shown in \ccite{vv}{kurchan}, if
one  counts the stationary points of a function neglecting the modulus
in  integrals of  the kind \re{mink}, a trivial result is obtained,
due to the Morse theorem.  Nevertheless, we note that we are not
calculating here the whole number of  stationary points of the TAP
free energy, but we are counting those  {\it with a given energy
density}. Moreover, for energies lower than the threshold, the
dominant contribution to the determinant is  truly positive, as can be
shown by computing the Hessian spectrum of the TAP free energy. On the
other hand, the same procedure of removing the modulus gives the
normal unconstrained TAP complexity of \cite{crisatap}, which has been
confirmed in \cite{franzparisi} with a totally different  method.

In order to perform the calculation it is useful to write both these
determinants by means of a Fermionic representation
$$
\det A = \int  d\bar\psi d\psi \ e^{- \sum_{ij}\bar\psi_i A_{ij} \psi_j}
\eq{det}
$$
while the usual (Bosonic) integral representation is adopted for the
delta functions which implement the TAP equations in \re{mink}.  The
average over the  disorder $J$ generates couplings among Bosonic and
Fermionic variables, but these mixed couplings are set equal to zero
as in the unconstrained  calculation. In this way it is possible to
average separately the Fermionic part from the Bosonic one.  This
simplifies a lot the calculation, since it turns out that the
Fermionic part  has exactly the same form as in the unconstrained
case, while, due to the  presence of the constraint $q_0$, this is no
longer true for the Bosonic part.  We can write the unconstrained
complexity in the usual following way (see 
\cite{crisatap}):
$$
\Sigma(E)=\Xi(E)_{Bosons}+\Omega(E)_{Fermions}
\eq{lupa}
$$
with
$$
\Xi(E)=\ {1\over 2}\log (2/p)-{1\over 2} - E^2  \quad ; \quad
\Omega(E)=-pEz-{p(p-1)\over 4}z^2-\log z   
\eq{xiome}
$$
$$
z={-E-\sqrt{E^2-2(p-1)/p}\over (p-1)} \ .
\eq{zeta}
$$
Similarly, for what said above, we can write
$$
\Gamma(E_1,E_2,q)= \Delta(E_1,E_2,q) + 
\Omega(E_1)+\Omega(E_2) 
\eq{guzzi}
$$
where $\Delta$ is the Bosonic contribution to \re{mink}. In this way
$\Sigma_c$ has the form
$$
\Sigma_c(E_2,q|E_1)= \Delta(E_1,E_2,q)  - 
\Xi(E_1)+ \Omega(E_2)  \ .
\eq{finz}
$$
Besides the variables $\sigma$ and $\tau$ there are two more Bosonic
fields coming from the integral representation of the delta functions,
respectively $\mu$ and $\lambda$.  All these Bosons couple because of
the average over the disorder $J$. To  perform the saddle point
approximation we introduce the following set of  variational
parameters:
$$
\eqalign{
Nx_1=\mu\cdot\mu                 \quad;\quad  Nx_2=& \ \lambda
\cdot\lambda         \quad;\quad Nx_3=\mu\cdot\lambda
\cr Ny_1=\sigma\cdot\mu               \quad;\quad
Ny_2=\tau\cdot\lambda               \quad; & \quad
Ny_3=\sigma\cdot\lambda               \quad;\quad Ny_4=\tau\cdot\mu
\ .          \cr}
\eq{para}
$$
We remind that $\sigma\cdot\tau=Nq_0$, while
$\sigma\cdot\sigma=\tau\cdot
\tau=N$. The explicit calculation gives
$$
\eqalign{ 
\Delta(E_1,E_2,q)=&                 \cr
=&\ \bbuildrel\hbox to .25in{\rm Ext}_{x,y} \bigg\{  \  ipE_1y_1 +
ipE_2y_2 - {p(p-1)\over 2}q_0^{p-2}y_3y_4 -{p\over 4}(x_1+x_2) -
{p\over2 }q_0^{p-1}x_3  \cr &\phantom{\ \bbuildrel\hbox to .23in{\rm
Ext}_{x,y} pe  } - {p(p-1)\over 4}(y_1^2+y_2^2)   +{1\over
2}\log\left({\Lambda\over 1-q_0^2}\right)  \bigg\}       \cr}
\eq{del}
$$
with
$$
\Lambda=[x_1(1-q_0^2)-k_1][x_2(1-q_0^2)-k_2]-[x_3(1-q_0^2)-k_3]^2
\eq{a}
$$
$$
k_1=y_1^2-2q_0y_1y_4+y_4^2 \ \ ; \ \ k_2=y_2^2-2q_0y_2y_3+y_3^2 \ \ ;
\ \  k_3=y_1y_3-q_0(y_1y_2+y_3y_4)+y_2y_4 
\eq{spata}
$$
and we remind that $q_0 = q / \sqrt{q_1 q_2}$.  The saddle point
equations are easily solved numerically, while it is possible  to
check analytically that  for $q_0=0$ we have
$\Delta(E_1,E_2,0)=\Xi(E_1)+
\Xi(E_2)$ and then equation \re{check} is fulfilled.


\autosez{arrriba} Normal and anomalous regimes.
\par

In this Section we analyze  the dependence of the constrained
complexity $\Sigma_c$ on $E_2$ and $q$, at a fixed value of $E_1$.
In particular, we focus our attention on the dependence of $\Sigma_c$ on
the overlap $q$, since we are interested in the overlap spectrum of 
the system, this giving information on the geometrical distribution of the
states in the phase space. In what follows it is always assumed  $T\in
[T_s,T_d]$, where $T_s$ and $T_d$ are respectively the static and the
dynamical transition temperatures. 

As a value for the reference energy $E_1$ we choose
$E_1=E_{eq}(\beta)$, where $E_{eq}(\beta)$  is the zero temperature
energy density of those TAP solutions that  dominate  at temperature
$\beta$; in this way, we are fixing an equilibrium state and  counting
how many states of energy density $E_2$ are present at distance $q$
from it. We stress however that we could choose any other value for
the reference energy $E_1$, obtaining qualitatively the same results. 

It is useful to distinguish two different regimes:  a {\it normal}
regime,  corresponding to values of $E_2$ well below the threshold, in
which geometrical considerations  at least qualitatively apply, and an
{\it anomalous} regime, characterized by values of $E_2$ just below
the threshold, which shows a rather peculiar  behaviour of $\Sigma_c$.

\vskip 1.0 truecm

\item{$\bullet$}{\it The normal regime:}

\vskip 0.2 truecm

The normal regime is defined by values of the energy $E_2$ of the
states we  are counting well below the threshold energy $E_{th}$.
Intuitively, we expect that $\Sigma_c$ decreases  with increasing $q$,
since this corresponds to consider smaller and smaller portions of the
phase space into which looking for TAP solutions. This is indeed what
happens in the normal regime, as shown in  Figure 1, where we have
plotted $\Sigma_c$ as a function of $q$.

\includegraphics{panta-fig1.ps}
\vbox{
\hbox{ \hglue   2.5 truecm \vbox{$ \Sigma_c$ \vglue 3.3 truecm}
\vbox{\vglue 6.7 truecm}
} \hbox{     \hglue 8 truecm $ q$ } \vbox{ \vglue 0.06 truecm} \vbox{
\hsize=16 truecm
\baselineskip=10 pt   

\piccolo{\noindent  {\pbf Figure 1}: 
The constrained complexity $\s \Sigma_c$ as a function of the overlap
$\s q$ (full line), with   $\s E_1=E_2=E_{eq}=-1.156$, for $\s
\beta=1.64$  and $\s p=3$. The self overlap of the two states is $\s
q_1=q_2=0.55$,  while  $\s \Sigma_c=0$ at $\s q_{last}=0.25$.  The
dashed-dotted line is the curve predicted by a random distribution of
the states (see the text).  For $\s q=0$ both the curves coincide with
the unconstrained complexity $\s \Sigma(E_2)$.  }} }
\vskip 0.5 truecm

This curve provides us some important information: 
First of all this is a {\it continuous} curve, meaning that there is a
continuous spectrum of states with energy $E_2$ around the fixed
reference state of energy $E_1$  (in Figure 1 we have set
$E_2=E_{eq}$).  This means that there is  an exponentially high number
of states at any value of $q$, until a value $q_{last}$ for which
$\Sigma_c=0$.  Thus $q_{last}$ gives the overlap of the nearest states
with energy $E_2$. It is important that, as long as $E_1$ and $E_2$
are different from $E_{th}$, this value $q_{last}$ is {\it smaller}
than the self overlap of the considered states. This gap between the
last value of the overlap in the spectrum and the self overlaps
$q_1,q_2$  simply means that these states are well separated one with
respect to the other, i.e. 
$$
{q_{last} \over \sqrt{q_1q_2}} < 1 \ .
\eq{usula}
$$
This has to be compared with the case of the Sherrington-Kirkpatrick
model
\cite{sk} in which as well as in this case there is a continuous distribution 
of states, but with mutual overlap going from $q=0$ up to the self
overlap  $q=q_{EA}$ \ccite{spin}{ultra}.

The second important feature of Figure 1 can be caught if we compare
$\Sigma_c$ with the corresponding quantity that would be obtained in
the case of a {\it random} distribution of the states. The simplest
hypothesis we can formulate on the geometrical structure of the states
is  that they are randomly distributed in the phase space: in this
case, the number of states at distance $q$ from a given fixed point in
the phase space would be simply given by the total number of states
multiplied by the volume of the manifold defined by fixing $q$, i.e.
$$
\Sigma_{random}=\Sigma(E_2)+{1\over 2}\log\left(1-{q^2\over q_1q_2}\right)
\eq{marna}
$$
where $\Sigma(E_2)$ is the unconstrained complexity and $q_1,q_2$ are
the self overlaps of the two solutions. If we plot this quantity as a
function of $q$ and compare it with $\Sigma_c(q)$ (Figure 1),  it can
be seen that there is a violation of the random distribution and that
this violation goes in the  direction of having an higher number of
states when looking at small distances.

\includegraphics{panta-fig2.ps}
\vbox{
\hbox{ \hglue   2.5 truecm \vbox{$ \Sigma_c$ \vglue 3.3 truecm}
\vbox{\vglue 6.7 truecm}
} \hbox{     \hglue 8 truecm $ q$ } \vbox{ \vglue 0.06 truecm} \vbox{
\hsize=16 truecm
\baselineskip=10 pt   

\piccolo{\noindent  {\pbf Figure 2}: 
The constrained complexity $\s \Sigma_c$ (full line),  compared with
the random distribution \re{marna}  (dashed-dotted line) for $\s
p=30$. All other parameters are the same as in Figure 1.  }} }
\vskip 0.5 truecm

The comparison with the random distribution of the states suggests an
interesting check. It is known that the $p$-spin model in the limit
$p\to\infty$ coincides with the Random Energy Model of \cite{rem},
which is characterized by a complete decorrelation of the energy
levels of the system. Therefore we expect that for increasing values of
$p$ the constrained complexity $\Sigma_c$ gets more and more similar
to the random distribution
\re{marna}. This is shown in Figure 2, where we see that for $p=30$ it remains
only a little tail for large $q$  in which the two distributions are
different.  For $p\to\infty$ they coincide.  We note that in the
Random Energy Model there is no geometrical structure at all. What we
have here is that a complete {\it energetic} decorrelation between
different states corresponds to a complete {\it geometrical}
decorrelation in the phase space.   

Finally we consider the dependence of $\Sigma_c$ on the energy $E_2$
of the states that we are counting. In Figure 3 we have plotted $\Sigma_c(E_2)$
at various values of $q$. The curve on the top corresponds to $q=0$ and then
reproduces the unconstrained complexity $\Sigma(E_2)$ (see equation 
\re{check}). We note that, even at fixed $q\neq 0$, $\Sigma_c$ increases 
with increasing $E_2$, as in the unconstrained case. 
Furthermore, as expected, the whole curve $\Sigma_c(E_2,q)$ decreases with 
increasing $q$.

\includegraphics{panta-fig3.ps}
\vbox{
\hbox{ \hglue   2.5 truecm \vbox{$ \Sigma_c$ \vglue 3.3 truecm}
\vbox{\vglue 6.7 truecm}
} \hbox{     \hglue 8 truecm $ E_2$ } \vbox{ \vglue 0.06 truecm}
\vbox{ \hsize=16 truecm
\baselineskip=10 pt   

\piccolo{\noindent  {\pbf Figure 3}: 
The constrained complexity $\s \Sigma_c$ as a function of the energy
$\s E_2$, at various values of the overlap $\s q$ and  $\s
E_1=E_{eq}=-1.156$, for  $\s \beta=1.64$ and $\s p=3$.  The first
curve on the top has $\s q=0$ and thus corresponds to the
unconstrained complexity $\s  \Sigma(E_2)$.  The threshold energy is
$\s E_{th}=-1.1547$.  The squares indicate the minimum of the function
$\s  \phi_c$ (see Section 4), which reaches the axis $\s \Sigma_c=0$ 
at $\s q_{low}=0.113$.       }} }
\vskip 0.5 truecm

\vskip 1.0 truecm 

\item{$\bullet$}{\it The anomalous regime:}

\vskip 0.2 truecm

We turn now to examine the anomalous regime.  If we plot $\Sigma_c$ as
a function of $E_2$ exactly as in Figure 3, but now expanding a narrow
range of energies $E_2$  just below the threshold ($E\in
[E_{th}-5\times 10^{-4},E_{th}]$), we obtain the behaviour shown in Figure 4.

The curves get lower and lower with increasing $q$, until
for a value of $q$ that we call $q_{max}$ they begin to reverse,
folding upwards. In Figure 4, $q_{max}$ corresponds to the curve whose
intersection with the $\Sigma_c=0$ axis starts going backward to the left 
with increasing $q$.
If now we make a section of these curves at a fixed value of $E_2$ in
this  range, plotting $\Sigma_c$ as a function of $q$, we find the
anomalous  behaviour of Figure 5: there is an interval of $q$ where
$\Sigma_c$  {\it increases} with increasing $q$.

\eject

\includegraphics{panta-fig4.ps}
\vbox{
\hbox{ \hglue   2.5 truecm \vbox{$ \Sigma_c$ \vglue 3.3 truecm}
\vbox{\vglue 6.7 truecm}
} \hbox{     \hglue 8 truecm $ E_2$ } \vbox{ \vglue 0.06 truecm}
\vbox{ \hsize=16 truecm
\baselineskip=10 pt   

\piccolo{\noindent  {\pbf Figure 4}: 
The constrained complexity $\s \Sigma_c$ as a function of the energy
$\s E_2$ just below the threshold, at various values of 
$\s q$. 
The value $\s q_{max}$ corresponds to the curve whose intersection with
the axis $\s \Sigma_c=0$ starts going to the left with 
increasing $\s q$.  }} }
\vskip 0.5 truecm

\includegraphics{panta-fig5.ps}
\vbox{
\hbox{ \hglue   2.5 truecm \vbox{$ \Sigma_c$ \vglue 3.3 truecm}
\vbox{\vglue 6.7 truecm}
} \hbox{     \hglue 8 truecm $ q$ } \vbox{ \vglue 0.06 truecm} \vbox{
\hsize=16 truecm
\baselineskip=10 pt   

\piccolo{\noindent  {\pbf Figure 5}: 
The constrained complexity $\s \Sigma_c$ as a function of the overlap
$\s q$ in the anomalous regime.  $\s E_1=E_{eq}$ and $\s E_2=-1.1550$.
}} }
\vskip 0.5 truecm

The reversed behaviour of $\Sigma_c$ clearly has not a geometrical
origin  and for this reason we talk of an anomalous regime. What it
seems is that, given a state (in this case an equilibrium state),
there is a sort of  clustering of states with high energy at small
distances (large overlap) from it, thus giving the distribution of
Figure 5.  Moreover, from Figure 4 we note that for high enough values
of $q$,  $\Sigma_c$ is no longer monotonic with respect to $E_2$, and
it develops a  maximum. Therefore, in this range of $q$ threshold
solutions are no longer  the most numerous around the fixed
equilibrium state.

To conclude, we note that the amplitude of the anomalous regime
depends on  the reference energy $E_1$: the range of energies $E_2$
into which  this anomalous behaviour occurs is larger for low values
of $E_1$ and shrinks to zero as $E_1$ approaches $E_{th}$.


\autosez{piz} The spectrum of the dominant states.
\par

At a first sight the increasing of $\Sigma_c$ with $q$, together with
the role of $q_{max}$ in the anomalous regime could seem an artifact
of the calculation. Nevertheless, how we are going to  show, this
behaviour is able to explain the energy spectra obtained with the real
replica method of \cite{noi}, where a completely different kind  of
computation is performed. 

To face this problem we have to ask: Which are the
{\it dominant} states at overlap $q$ from a given fixed state ?
In the unconstrained case \cccite{kpz}{crisatap}{buribarrameza},
equilibrium states of the system are defined as those TAP solutions
that minimize the generalized free energy density $\phi(f)=f-
T\Sigma(f)$.  Similarly, we can wonder which  solutions dominate at
distance $q$ from a given equilibrium state.  To this end we look for
the minimum of the function $\phi_c(f,q)=f- T\Sigma_c(f,q)$, with
$E_1=E_{eq}$ and $\Sigma_c$ expressed as a function of the free energy
density $f$ of the states found at distance $q$.  In Figure 3, for
each given value of $q$, we have signed  on the corresponding curve
the point that minimizes $\phi_c$.  As it is easily seen, there is a
value of $q$ for which the minimum of  $\phi_c$ reaches the axis
$\Sigma_c=0$.  We call  this value $q_{low}$. If we look for a minimum
of $\phi_c$ when  $q>q_{low}$  we would be brought to a negative value
of $\Sigma_c$, i.e.  to non existing solutions (in the limit
$N\to\infty$).  In this situation, the dominant  solutions are those
with the lowest energy density and a non negative value of the
complexity, i.e. with $\Sigma_c=0$.  Due to this, at $q_{low}$ there
is a sharp change in the energy behaviour of the dominant solutions:
this  energy  decreases following the minimum of $\phi_c$ until
$q=q_{low}$, then,  for higher values of $q$, it increases following
the intersection of the curves  $\Sigma_c(E_2)$ with the axis
$\Sigma_c=0$. Yet, from Figure 4 we see that when $q$ reaches
$q_{max}$, due to the anomalous behaviour of $\Sigma_c$ this
intersection  point inverts its direction, and so the energy of the
dominant solutions starts to decrease. 

We conclude that the energy density of the solutions dominating at
distance $q$ from a fixed equilibrium state has a cuspid for
$q=q_{low}$ and reaches a  maximum at $q=q_{max}$. What said above is
shown in Figure 6.

\includegraphics{panta-fig6.ps}
\vbox{
\hbox{ \hglue   2.5 truecm \vbox{$ E_2$ \vglue 3.3 truecm}
\vbox{\vglue 6.7 truecm}
} \hbox{     \hglue 8 truecm $ q$ } \vbox{ \vglue 0.06 truecm} \vbox{
\hsize=16 truecm
\baselineskip=10 pt   

\piccolo{\noindent  {\pbf Figure 6}: 
The energy density $\s E_2$ of the solutions dominating at  distance
$\s q$ from the fixed equilibrium solution.  The cuspid is at $\s
q_{low}=0.113$ and the maximum is at $\s q_{max}=0.360$. The last
point of  the curve represents the distance of the nearest solutions.
}} }
\vskip 0.5 truecm

In the real replica method usually a first replica is quenched into an
equilibrium state, while a second replica is forced to thermalize at
distance  $q$ from it \ccite{franzparisi}{noi}.  It is then natural to
think that the second replica thermalizes  into one of the states that
dominate at distance $q$ in the sense   described above, and that
therefore the spectrum of the states visited by this second  replica
is of the same kind as the spectrum of Figure 6.  This is indeed what
happens, as  shown in more details in Appendix A: the spectrum found
with the real replica method of \cite{noi} presents a cuspid, and 
has a maximum exactly at $q=q_{max}$, thus providing a  confirmation 
of the existence of the anomalous regime. 

In analyzing Figure 6 it is important to note that the number of
dominant states is  exponentially high in $N$ as long as $q<q_{low}$,
while it is of order $N$  for $q>q_{low}$. In terms of constrained
systems this transition is signaled by the breaking of the replica
symmetry in the overlap matrix and this is  another confirmation of
the consistency of these two different methods   (see Appendix A)
\ccite{noi}{bfp}.

The ending point of this curve corresponds to that value of the
overlap  above which no states of any energy are found with $\Sigma_c
\geq 0$. Thus it  indicates which  is the overlap with the fixed state
of energy $E_1=E_{eq}$ of the  states nearest to it. It turns out that
these nearest states have an energy density {\it greater} than the
energy $E_{eq}$ of the fixed reference state, even if in general
they are not threshold states.  What is important is that  this is
not a special feature of the equilibrium  states: indeed, given a
state of {\it any} energy $E_1$, the states nearest to it {\it
always} have an energy density greater than $E_1$. 

This behaviour is not a trivial consequence of the fact that states with 
higher energy are in general more numerous, since following this reasoning 
the nearest states would always be the threshold ones (which have the greatest
unconstrained complexity), while we know that this is not true. 
Indeed, as mentioned in Section 3, at large values of the overlap $q$, i.e 
in the anomalous regime, threshold states are no longer the most numerous 
around a given fixed state.
Once the curve of the dominant states of Figure 6 develops a maximum at 
$q_{max}$, it is not obvious which should be the energy of the states 
corresponding to the ending point of this curve.


\autosez{soglia} The threshold states.
\par

We turn now to examine the structure of the  threshold states. We
remind that threshold states are those  solutions of the TAP equations
with the highest energy density $E_{th}$. These states have a great
importance under several aspects. 

From a static point of view it can be shown that threshold states are
{\it marginal}:  the typical  spectrum of the free energy Hessian
evaluated in a TAP solution of energy $E$  is a semicircle with
support in the positive semi axis and its lowest  eigenvalue
$\lambda_{min}$ goes to zero as $E$ goes to $E_{th}$. In this  sense
threshold states develop some flat directions.    

On the other hand, in the temperature range $T<T_d$, threshold
states play a fundamental role in the off-equilibrium dynamics of
this model.   Solving analytically the dynamical equations with random
initial  conditions (i.e. high energy initial conditions), one finds
that the  asymptotic limit  of the energy ${\cal E}_{\infty}$
coincides with the threshold energy  $E_{th}$. In other words, the
asymptotic dynamics takes place  at the threshold level, never
visiting the sub threshold landscape. Moreover,  the dynamical
evolution of  the system presents a first  equilibrium regime in which
the correlation function goes to the value of  the self overlap of the
threshold states $q_{th}$, followed by an  off-equilibrium  aging
regime in which the correlation function goes to zero,  this meaning
that the system never truly thermalizes into any of the threshold
states, but that rather it continuously drifts away  \cite{ck1}. 

Therefore it is intriguing to investigate the eventual relations
between the peculiar dynamical behaviour that occurs at the threshold
level and the geometrical structure of the threshold states.

To perform an analysis of this kind we set $E_1=E_2=E_{th}$, and we
study the constrained complexity $\Sigma_c$ as a function of $q$, i.e.
we fix a threshold solution and count how many other threshold
solutions are present at distance $q$ from it. The corresponding curve
is shown in Figure 7.

The important result is that
$$
\Sigma_c(q)\to 0 \quad {\rm for} \quad q\rightarrow q_{th}(\beta)
\eq{ulla}
$$
where $q_{th}(\beta)$ is the self overlap of the threshold states at
the temperature $\beta$ we are considering.  More precisely we find
$$
\Sigma_c(q)\sim (q_{th}-q)^5 \quad {\rm for} \quad q\sim q_{th} \ .
\eq{cinque}
$$

\includegraphics{panta-fig7.ps}
\vbox{
\hbox{ \hglue   2.5 truecm \vbox{$ \Sigma_c$ \vglue 3.3 truecm}
\vbox{\vglue 6.7 truecm}
} \hbox{     \hglue 8 truecm $ q$ } \vbox{ \vglue 0.06 truecm} \vbox{
\hsize=16 truecm
\baselineskip=10 pt   

\piccolo{\noindent  {\pbf Figure 7}: 
The constrained complexity $\s \Sigma_c$ as a function of  $\s q$ with
$\s E_1=E_2=E_{th}$, for $\s \beta=1.64$ and $\s p=3$.  For $\s
q>0.35$ the curve is indistinguishable from the axis and it reaches
zero for $\s q=q_{th}=0.504$.  }} }
\vskip 0.5 truecm

When the overlap between two states of the same kind is equal to their
self overlap, it means that these two states are coincident in the 
thermodynamic limit. 
Thus, thinking of well separated states, one expects that  $\Sigma_c$ goes
to zero at a value of $q$ which is  lower than the self overlap:
indeed this is what happens for all the states below the threshold
(see Figure 1 and equation \re{usula}).

On the other hand, from \re{ulla} we see that, fixed a threshold
state,  other threshold states with vanishing complexity are found until
a distance zero from the fixed one.
A similar conclusion was deduced in \cite{noi}, but in that context it
was possible to state this result only at the dynamical transition
temperature $\beta=\beta_d$, at which threshold states are the
equilibrium ones. Here we see that this remains true at any
temperature,  as a natural consequence of the non chaoticity of TAP
solutions with  temperature.

As a consequence of equation \re{ulla} we can say that there is no 
sharp separation among threshold states, and that they rather 
form a structure of coalescent states. This means that these states are 
separated by free energy {\it density} barriers which are vanishing in the
limit $N\to\infty$, i.e. that the free energy barriers grow as $N^\alpha$
with $\alpha<1$ \cite{lalu}. 
  
As previously said, these states are minima of the TAP free energy  
with some flat directions.
We can then argue that they are connected along these flat directions, 
forming a sort of channel of states. 
Into this frame the dynamical drifting of the system (i.e. the
decreasing to zero of the correlation function in the off-equilibrium
regime) can be viewed as a wandering along this channel of
threshold states, in agreement with the ideas outlined in 
\cite{ck1}.

Finally, we note that, due to equation \re{ulla}, the distribution of
threshold states has no gap between the last value of the overlap in
the spectrum and the self overlap, unlike all other sub threshold
states. This feature is reminiscent of the overlap distribution in
the Sherrington-Kirkpatrick model. Besides, we remind that in the SK
model all the states are marginal, as it happens for the threshold
states in the  $p$-spin spherical model. From this point of view, we
can say that threshold states are the most SK like.


\autosez{cc} Conclusions and open questions.
\par

The $p$-spin spherical model for $T_s<T<T_d$ is dominated 
by an exponentially high number of states, each one having a rather
high free energy density $f$ and therefore very small weight. This high
free energy is counterbalanced by the entropic contribution of the complexity.
Besides these
equilibrium states, there is a great variety of metastable states, with free 
energy densities both lower and larger than the equilibrium one, all having
a vanishing weight. In this sense, there is no real difference between 
equilibrium and metastable states, being the equilibrium a collective property.

In this situation, the standard replica approach fails, since it is not able 
to discriminate the overlap relations among all these states, while the
TAP approach gives no information on the overlap distribution of them.

We have faced this problem introducing an overlap $q$ between different TAP
solutions and performing a purely entropic computation of their number. In 
this way we have defined a constrained complexity $\Sigma_c(q)$, which plays 
the role of the overlap distribution of the states.

By means of $\Sigma_c$ we have found that the states are disposed in 
a non trivial way:
fixed an arbitrary state from which observing the phase space, 
there is a continuous spectrum of states surrounding it. 
This means that there are states at each value of the 
overlap $q$ with the reference one, from $q=0$ until a maximum value 
$q_{last}$. For sub threshold states there is a gap between 
this last value of the overlap and the value of the self overlap, 
meaning that they are well separated states.
Moreover, the distribution is different from the one 
obtained supposing a random distribution of the states, since it shows
a major crowding of states at small distances. Yet, the two distributions 
coincide in the $p\to\infty$ limit, when the case of the Random Energy 
Model is recovered. 

Furthermore, the analysis of threshold states has given some interesting
results: these marginal states have an overlap distribution 
which goes continuously to zero at a value of $q$ equal to their 
self overlap $q_{th}$. 
This means that these states are connected along 
their flat directions, forming a sort of channel that winds along the 
phase space. This feature may have a role in the asymptotic dynamics of the 
system, which occurs at the level of the threshold landscape.

To comment the continuous distribution of the overlap, and in particular
the coalescent structure of the threshold states, we have often referred
to the SK model, for which a continuous structure with respect to
the overlap $q$ is directly given by the standard distribution function 
$P(q)$. 
However, in doing this comparison it is necessary to make some  
specifications. In the ultrametric structure of the SK model
the main role is played by the equilibrium states, which have the lowest free 
energy and finite weight, and whose number is of order $N$. This 
means that one can disregard the exponentially high number of
states with higher free energy and vanishing weight 
(this can be done by introducing a cut-off on the branches size of 
the ultrametric tree \cite{ruelle}). The situation for the $p$-spin spherical
model is somehow the opposite: here the states, both of equilibrium and 
metastable, {\it all} have vanishing weight and are in exponentially high 
number. Therefore none of these states, whatever energy it has, can be 
disregarded. 
In this sense the investigation of the structure of the states 
for this model is complicated by the existence of a new relevant variable, 
that is the energy. 

Moreover, we stress that there is at the present moment no evidence of
an ultrametric organization of the states in the $p$-spin spherical
model. To investigate this point it would be necessary to analyze the 
correlation among triplets of states, as it has been done for the SK 
model in \cite{ultra}. The only fair indication of a clustering structure 
of the states comes from the existence of the anomalous regime, into which
the constrained complexity grows with the overlap $q$. The existence
of this anomalous regime is confirmed by the real replica method.
   
By means of the constrained complexity we have also obtained the spectrum
of the dominant states at distance $q$ from a given fixed state. 
This spectrum shows that below a certain value
of the overlap, the number of the dominating states is exponential in $N$, 
while above this value (close states) this number is of order $N$. Besides,
we have seen that the states nearest to any given state always have higher
energy density.

We want to mention here a working hypothesis that could be useful
for the present investigation.
In the Generalized Random Energy Model of \cite{grem}, an 
ultrametric structure is directly built in by defining the
probability distributions of the energies at each clustering level.
In this context, an equivalent of the functional order parameter $x(q)$ 
of \ccite{rsb1}{rsb2} can be identified.
In \cite{grem} this construction is explicitly performed in the case
of two clustering levels: in this simple example one can see
that a function $x(q)$ which {\it decreases} with increasing $q$ 
corresponds to a hidden ultrametric
structure, in the sense that this structure is present by construction, but
is not revealed from the computation of the free energy of the system. 
This suggests that also for the $p$-spin spherical model, 
where a rich distribution of states is present but hard to reveal, 
an anomalous function $x(q)$ could describe the underlying hidden structure.  
The existence of anomalous solutions of this kind in the context of the
replica approach has been shown and discussed in 
\ccccite{kpz}{ferro}{bava}{vira}.

To conclude, the main open issue on this topic is the pursuit of a 
unifying frame into which inserting all the results we have obtained, in 
order to describe in a synthetic way the rich and complex structure of states 
of the $p$-spin spherical model.


\vskip 2 truecm
\noindent
{\bf Acknowledgements.}

It is a pleasure to thank for important suggestions and 
very useful discussions Alain Barrat, Leticia Cugliandolo, Silvio Franz,
Jorge Kurchan, Marc M\'ezard, R\'emi Monasson, Heiko Rieger, Felix Ritort 
and Miguel Angel Virasoro.

\eject


\vskip 2 truecm 
\semiautosez{\rm A} Appendix A: The comparison with the real replica method.
\par

The real replica method \ccccite{kpz}{franzparisi}{noi}{frapavi}
consists in studying the static properties of a certain number of real 
replicas of the system, as a function of the overlaps imposed among them.
In \cite{noi} we introduced a three replica potential by which we analyzed 
the structure of equilibrium and metastable states of the 
$p$-spin spherical model: the first replica is fixed into an equilibrium 
state, while 
the second replica is forced to equilibrate at overlap $q_{12}$ with the 
first one. Clearly, the way in which replica 2 chooses its 
constrained equilibrium state is heavily conditioned by the complexity 
$\Sigma_c$ of the states at that distance. Once fixed replicas 1 and 2, the
potential $V_3$ is defined as the free energy density 
of a third replica 3 as a 
function 
of its distances $q_{13}$ and $q_{23}$ from the first two. 
The most important
minimum of $V_3$ corresponds to replica 3 in local equilibrium into 
the state chosen by replica 2. In this minimum, that we call $M_2$,  the 
energy and the  self overlap of  replica 3  satisfy TAP equations, meaning
that replicas 2 and 3 have found a state of the unconstrained system at 
distance $q_{12}$ from the equilibrium state of replica 1. 

Before proceeding further it is crucial to clarify the different roles of 
the three replicas, beginning from the way in which replica 2 chooses the state 
at distance $q_{12}$ from 1. If the TAP free energy is minimized {\it on} the 
manifold defined by fixing $q_{12}$, the number of minima found is clearly  
greater than the number of {\it genuine} TAP solutions at that distance and
for the most part consists of projections on the manifold of nearby true
TAP solutions. These projections are what replica 2 sees as states. 
Therefore, we can say that replica 2 thermalizes in the vicinity  
of a TAP solution, but this solution is in general at distance $q\neq q_{12}$
from replica 1. On the other hand, this TAP solution is that into which 
replica 3 thermalizes giving the minimum $M_2$, and this is why replica 3 
gives the right TAP energy of this state and its right distance from replica 1,
while replica 2 does not. To understand the energy spectrum of the states
visited by replica 3, we must then investigate how replica 2 chooses these 
states.

As usual, replica 2 tries to optimize the balance between the free energy 
and the complexity of the solutions it finds on the manifold, minimizing a 
function $\phi_2=f_2-T\Sigma_2(f_2,q_{12})$. It is important to note
that $\Sigma_2$ is {\it not} the same function as $\Sigma_c$: 
\noindent
There are {\it many} genuine TAP solutions with different TAP free energies 
$f$ and at 
slightly different distances $q$, either higher or lower than $q_{12}$, whose
projections on the manifold of replica 2 all have the same free energy $f_2$.
Of these TAP solutions the relevant ones are those with the maximum complexity
$\Sigma_c(f,q)$. This maximum value will then give $\Sigma_2(f_2,q_{12})$.
Summarizing, $\Sigma_2(f_2,q_{12})$ is equal to the complexity $\Sigma_c$
of the most numerous TAP solutions whose projections on the manifold fixed by 
$q_{12}$ have free energy $f_2$ (the computation of $\Sigma_2$ is explicitly 
performed in \cite{bfp}).

The minimization of $\phi_2$ with respect to $f_2$ then selects a 
particular class of TAP solutions. Replica 2 is quenched into one of these
solutions in the distorted way explained above, while replica 3 truly 
thermalizes into it. 
For this reason, the dependence on $q_{12}$ of the energy of replica 3 
in the minimum $M_2$ is a mere 
manifestation of the process of equilibration of replica 2 above described.

Resuming, the states visited by replica 3 are chosen minimizing $\phi_2$, 
which does not coincide with the function $\phi_c$ of Section 4 as well as 
$\Sigma_2$ is not the same function as $\Sigma_c$. Nevertheless, it is clear
that the spectrum given by the minimization of $\phi_2$ must be similar to
the one given by $\phi_c$ in Section 4. 

To make then a comparison with Figure 6, it is convenient to plot the zero
temperature energy of replica 3 in the minimum $M_2$ as a function of 
the overlap $q_{13}$, 
instead of $q_{12}$, since, as stated above, $q_{13}$ better represents
the real overlap between the two states. This parameterization is possible
because in the minimum $M_2$ the values of $q_{13}$ is uniquely determined 
by $q_{12}$.

\eject 

\includegraphics{panta-figa1.ps}
\vbox{
      \hbox{ \hglue   2.5 truecm \vbox{$ E$ \vglue 3.3 truecm}
                 \vbox{\vglue 6.7 truecm}                             }
      \hbox{     \hglue 8 truecm $ q_{13}$                                 }
      \vbox{ \vglue 0.06 truecm}
      \vbox{ \hsize=16 truecm  \baselineskip=10 pt   

\piccolo{\noindent  {\pbf Figure A1}: 
                    The energy density $\s E$ of replica 3 in the 
                    minimum $\s M_2$
 		    of the three replica potential, as a function of the
		    overlap $\s q_{13}$, for $\s \beta=1.64$ and $\s p=3$. 
                    Here, as in Figure 6, $\s q_{max}=
		    0.360$.
                                    }}
							              }
\vskip 0.5 truecm

It can be seen from a comparison between Figure 6 and Figure A1 that, as
we expected, the processes of minimization of $\phi_c$ and $\phi_2$ are 
qualitatively the same:

Firstly, we note from Figure A1 that there is a value of $q_{13}$ 
for which the curve has a cuspid; we call the corresponding value  
of $q_{12}$ in the minimum $M_2$ for which this cuspid occurs, $q_{rsb}$.
Indeed, following the same line of reasoning of Section 4, we 
argue that this cuspid must correspond to the point in  which $\Sigma_2$ 
becomes zero. The computation developed in \ccite{noi}{bfp} shows that for 
value of $q_{12}$ greater than $q_{rsb}$ the overlap matrix $Q^{22}$ of 
replica 2 breaks the replica symmetry. 
Physically, an RS form of the overlap matrix means either that
the systems finds an exponentially high number of states (as in the $p$-spin
model for $T_s<T<T_d$), or that the phase space consists in just one state
(as in the paramagnetic case). On the other hand, an RSB form means that 
the phase space is dominated by a number of order $N$ of states (as in the
$p$-spin model for $T<T_s$). Therefore, the breaking of the replica symmetry
of $Q^{22}$ for $q_{12}=q_{rsb}$ is a strong indication that here replica 
2 ceases to see an exponentially high number of dominant states and confirms 
that in this point $\Sigma_2$ becomes zero, as we argued above. 
As already stressed, the functions $\phi_2$ and $\phi_c$, even though 
they have a similar physical meaning, are actually different and this is 
why the corresponding minimization curves (first branches in Figures 6 and A1) 
are different.

Secondly, in Figure A1 we note a maximum for $q_{13}=q_{max}$, i.e. exactly 
in the same point as in Figure 6. 
This is a consequence of the reversing in the behaviour
of $\Sigma_c$ that occurs in the anomalous regime. This behaviour is directly 
inherited by $\Sigma_2$: 
loosely speaking, if there are TAP solutions whose number starts to increase 
getting closer to the state of replica 1, the corresponding projections on a 
fixed manifold will increase too. This is an important confirmation of
the role of the quantity $q_{max}$ and, as a consequence, it is a confirmation
of the existence of the anomalous regime. 

Finally, it is worth  observing that the ending points of the two curves are 
different, i.e. the potential $V_3$ ceases to see states around replica 1 at 
a distance that is {\it greater} than the one corresponding to the nearest
TAP solutions given by $\Sigma_c$. This can signify either that the potential
fails to see these last states because replica 2 does not thermalizes in the
vicinity of them, or that these last solutions given by $\Sigma_c$
actually are not minima of the TAP free energy. At the present moment this 
is an open question.

Summarizing the considerations of this Appendix we can say that the
energy spectrum given by the real replica method is well explained by the
behaviour of the constrained complexity. This mutually confirms the
results found with the two methods. 


\vskip 3 truecm
\noindent
{\bf References.}
\vskip .5 truecm


\biblitem{ea} Edwards S F and Anderson P W 1975 {\it J. Phys. F: Metal. Phys.} 
{\bf 5} 965

\biblitem{sk} Sherrington D and Kirkpatrick S 1975 {\it Phys. Rev. Lett.} 
{\bf 35} 1792

\biblitem{tap} Thouless D J, Anderson P W and Palmer R G 1977 {\it Philos.
Mag.} {\bf 35} 593

\biblitem{rsb1} Parisi G 1979 {\it Phys. Rev. Lett.} {\bf 23} 1754 
                          
\biblitem{rsb2} Parisi G 1980 {\it J. Phys. A: Math. Gen.} {\bf 13} L115 
                                           
\biblitem{rsb3} Parisi G 1980 {\it J. Phys. A: Math. Gen.} {\bf 13} 1887

\biblitem{sompozip} Sompolinsky H and Zippelius A 1982 {\it Phys. Rev.} B 
{\bf 25} 6860

\biblitem{ck1} Cugliandolo L F and Kurchan J 1993 {\it Phys. Rev. Lett.}
{\bf 71} 173

\biblitem{tirumma} Kirkpatrick T R and Thirumalai D 1987 {\it Phys. Rev.} B 
{\bf 36} 5388

\biblitem{crisa1} Crisanti A and Sommers H-J 1992 {\it Z. Phys.} B 
{\bf 87} 341

\biblitem{crisa2} Crisanti A, Horner H and Sommers H-J 1993 {\it Z. Phys.} B
{\bf 92} 257

\biblitem{ck2} Cugliandolo L F and  Kurchan J 1994 {\it J. Phys. A: Math. Gen.} 
{\bf 27} 5749

\biblitem{kpz} Kurchan J, Parisi G and Virasoro M A 1993 {\it J. Phys. I France}
{\bf 3} 1819

\biblitem{franzparisi} Franz S and Parisi G 1995 {\it J. Phys. I France}
{\bf 5} 1401

\biblitem{ferrero} Ferrero M E and Virasoro M A 1994 {\it J. Phys. I France }
{\bf 4} 1819

\biblitem{buribarrameza} Barrat A, Burioni R and M\'ezard M 1996 
{\it J. Phys. A: Math. Gen}
{\bf 29} L81

\biblitem{monasson} Monasson R 1995 {\it Phys. Rev. Lett.}
{\bf 75} 2847

\biblitem{crisatap} Crisanti A and Sommers H-J 1995 {\it J. Phys. I France}
{\bf 5} 805

\biblitem{mezpa} M\'ezard M and Parisi G 1990 {\it J. Phys. A: Math. Gen.}
{\bf 23} L1229; M\'ezard M and Parisi G 1991 {\it J. Phys. I France } 
{\bf 1}  809

\biblitem{nieu} Nieuwenhuizen Th M 1996 {\it Phys. Rev. Lett. }
{\bf 74} 4289

\biblitem{franz} Franz S, private communication

\biblitem{vira} Virasoro M A 1996 {\sl Simulated annealing methods under 
analytical control}, in B-L Hao (Ed), ``Statphys 19'' - Proc. of the 19th 
IUPAP Int'l Conf. on Stat. Phys., Xiamen, China, World Scientific Pu\-bli\-
shing, Sin\-ga\-po\-re

\biblitem{grome} Gross D J and M\'ezard M 1984 {\it Nucl. Phys. B }
{\bf 240} 431

\biblitem{gard} Gardner E {\it Nucl. Phys. B}
{\bf 257} 747

\biblitem{noi} Cavagna A, Giardina I and Parisi G 1996 cond-mat 9611068, to
be published in {\it J. Phys. A: Math. Gen.}

\biblitem{noi2} Cavagna A, Giardina I and Parisi G 1997 cond-mat 9702069

\biblitem{braymoore} Bray A J and Moore M A 1980 {\it J. Phys. C: Solid. St. 
Phys.} {\bf 13} L469

\biblitem{frapavi} Franz S, Parisi G and Virasoro M A 1992 {\it J. Phys. I
France} {\bf 2} 1869
 
\biblitem{kurchan} Kurchan J 1991 {\it J. Phys. A: Math. Gen}
{\bf 24} 4969

\biblitem{vv} Vertechi D and Virasoro M A 1989 {\it J. Physique}
{\bf 50} 2325

\biblitem{bfp} Barrat A, Franz S and Parisi G 1997 cond-mat 9703091

\biblitem{rem} Derrida B 1981 {\it Phys. Rev.} {\bf B 24} 2613

\biblitem{ultra} M\'ezard M, Parisi G, Sourlas N, Toulouse G and Virasoro M A
1984 {\it J. Physique} {\bf 45} 843

\biblitem{spin} M\'ezard M, Parisi G and Virasoro M A 1987 {\sl Spin Glass 
Theory And Beyond}, World Scientific Pu\-bli\-shing, Singapore  
 
\biblitem{ruelle} Ruelle D 1987 {\it Commun. Math. Phys.} {\bf 108} 225

\biblitem{grem} Derrida B 1985 {\it J. Physique} {\bf 46} L401

\biblitem{prep} Cavagna A, Giardina I and Parisi G, in preparation

\biblitem{lalu} Kurchan J and Laloux L 1996 {\it J. Phys. A: Math. Gen}
{\bf 29} 1929

\biblitem{bava} Baviera R 1995 Tesi di Laurea, Universit\'a degli Studi di 
Roma ``La Sapienza'' 

\biblitem{ferro} Ferrero M E 1993 Tesi di Laurea, Universit\'a degli Studi di
Roma ``La Sapienza''

\insertbibliografia

\bye